%% file: main.tex
\documentclass[10pt,conference]{IEEEtran}

\usepackage{times}
\usepackage{soul}
\usepackage{url}
\usepackage[utf8]{inputenc}
\usepackage{caption}

\usepackage{amsmath,amssymb,amsfonts}
\usepackage{algorithm}
\usepackage{algorithmic}
\usepackage{graphicx}
\usepackage{textcomp}

\usepackage{booktabs} 
\usepackage{siunitx}
\usepackage{amsmath,xcolor,pifont}
\usepackage{xspace,url}

\usepackage{color}
\usepackage{colortbl}
\usepackage{multirow}
\usepackage{multicol}
\usepackage{makecell}
\usepackage{tabularx}
\usepackage{listings}
\usepackage{graphicx}
\usepackage{caption}
\usepackage{enumitem}
\usepackage{fancyvrb}
\usepackage{subcaption}

\usepackage[colorinlistoftodos]{todonotes}

\newtheorem{definition}{Definition}

\newcommand{\sys}{DeepPayload\xspace} 

\newcommand{\etal}{\textit{et al}.~}
\newcommand{\ie}{\textit{i}.\textit{e}.~}
\newcommand{\eg}{\textit{e}.\textit{g}.~}

\makeatletter
\newcommand{\linebreakand}{%
  \end{@IEEEauthorhalign}
  \hfill\mbox{}\par
  \mbox{}\hfill\begin{@IEEEauthorhalign}
}
\makeatother

\begin{document}
\title{\sys: Black-box Backdoor Attack on Deep Learning Models through Neural Payload Injection}

\author{
\IEEEauthorblockN{Yuanchun Li}
\IEEEauthorblockA{
Microsoft Research\\
Beijing, China\\
Yuanchun.Li@microsoft.com}
\and
\IEEEauthorblockN{Jiayi Hua}
\IEEEauthorblockA{
Beijing University of Posts and\\ Telecommunications,
Beijing, China\\
huajiayi@bupt.edu.cn}
\and
\IEEEauthorblockN{Haoyu Wang}\thanks{Corresponding author}
\IEEEauthorblockA{
Beijing University of Posts and\\ Telecommunications,
Beijing, China\\
haoyuwang@bupt.edu.cn}
\linebreakand
\IEEEauthorblockN{Chunyang Chen}
\IEEEauthorblockA{
Monash University\\
Melbourne, Australia\\
Chunyang.Chen@monash.edu}
\and
\IEEEauthorblockN{Yunxin Liu}
\IEEEauthorblockA{
Microsoft Research\\
Beijing, China\\
Yunxin.Liu@microsoft.com}
}

\maketitle

\begin{abstract}

Deep learning models are increasingly used in mobile applications as critical components. Unlike the program bytecode whose vulnerabilities and threats have been widely-discussed, whether and how the deep learning models deployed in the applications can be compromised are not well-understood since neural networks are usually viewed as a black box.
In this paper, we introduce a highly practical backdoor attack achieved with a set of reverse-engineering techniques over compiled deep learning models. The core of the attack is a neural conditional branch constructed with a trigger detector and several operators and injected into the victim model as a malicious payload.
The attack is effective as the conditional logic can be flexibly customized by the attacker, and scalable as it does not require any prior knowledge from the original model.
We evaluated the attack effectiveness using 5 state-of-the-art deep learning models and real-world samples collected from 30 users. The results demonstrated that the injected backdoor can be triggered with a success rate of 93.5\%, while only brought less than 2ms latency overhead and no more than 1.4\% accuracy decrease.
We further conducted an empirical study on real-world mobile deep learning apps collected from Google Play. We found 54 apps that were vulnerable to our attack, including popular and security-critical ones.
The results call for the awareness of deep learning application developers and auditors to enhance the protection of deployed models.
\end{abstract}







\input{sec_introduction}

\input{sec_background}
\input{sec_approach}

\input{sec_evaluation}

\input{sec_mitigation}

\input{sec_conclusion}

\section*{Acknowledgment}

We thank all anonymous reviewers for the valuable comments. Thank all volunteers who provided photos for real-world evaluation. Haoyu Wang is the corresponding author.

\bibliographystyle{IEEEtran}
\bibliography{reference}

\end{document}

%% file: sec_introduction.tex
\section{Introduction}


Deep neural networks (DNNs) have been used on a variety of tasks, including computer vision, natural language processing, recommendation systems, and medical diagnosis, where they have produced results comparable to and in some cases superior to human experts.
Due to the remarkable performance, DNNs have also been widely used in many security-critical applications, ranging from driving assistance \cite{wei2019enhanced}, user modeling \cite{li2018automated}, to face recognition \cite{sun2016sparsifying}, and video surveillance \cite{chung2017two}.
In these applications, DNN models are compiled and deployed to edge devices, such as mobile phones \cite{xu2019first,zhang2019DL_mobile_survey}, embedded vehicular systems \cite{hochstetler2018embedded}, and smart cameras \cite{ananthanarayanan2017real}. Various approaches have been introduced to optimize the performance \cite{he2018amc,lane2016deepx,zhang2019dabnn}, protect user privacy \cite{li2017privacystreams,liu2020pmc}, and secure the execution \cite{tramer2018slalom,lee2019occlumency,zhang2020dynamic} of these deployed models.

Despite the great advances, DNNs have been found vulnerable to various types of attacks~\cite{szegedy2013intriguing,trojannn}.
Backdoor attack (or Trojan attack) is one of the major attacks, which modifies the victim model to inject a backdoor (\ie a hidden logic). The backdoored model would behave as normal in most times, while producing unexpected behavior if a certain trigger is in the input.
For example,
a backdoored driving assistance model may give wrong prompts if an attacker-defined sign appears on the road. 
Unlike the adversarial attack that is widely-known as a robustness issue and many methods have been proposed to test \cite{pei2017deepxplore,tian2018deeptest,ma2018deepgauge, huang2021robustness}, enhance \cite{feng2020deepgini}, or verify \cite{paulsen2020reludiff} the robustness, the potential influence of backdoor attacks is not well understood.

The most representative approaches to achieve backdoor attacks are BadNets \cite{gu2017badnets,gu2019badnets} and TrojanNN \cite{trojannn}.
BadNets \cite{gu2017badnets} trains a backdoor into the DNN by poisoning the training dataset (\ie inserting a lot of adversarial training samples with the trigger sign).
TrojanNN \cite{trojannn} does not rely on access to the original training data. Instead, it extracts a trigger from the model and generates a small set of training samples that would change the response of specific neurons.

However, existing backdoor attacks can hardly be applied to published or deployed mobile/edge deep learning applications (DL apps for short), which are accessible to most adversaries.
First, both BadNets \cite{gu2017badnets} and TrojanNN \cite{trojannn} require training the victim model, which makes them inapplicable to deployed models where the parameters are frozen and optimized for inference \cite{tflite}.
Second, the triggers in their approaches are not practical for mobile/edge applications where the input images are directly captured by cameras. TrojanNN's triggers are irregular pixels computed from the model, which is hard or even impossible to generate in the physical world. BadNets supports arbitrary trigger, but it requires poisoning the training data before model training.
Thus, it remains unclear whether and how a post-training DL model can be compromised.

In this paper, we introduce \sys, a simple but effective black-box backdoor attack against deployed DNN models. Instead of training a backdoor into the model, \sys directly injects the malicious logic into the deployed model through reverse-engineering. We first disassemble the DNN model binary file to a data-flow graph, and then insert a malicious payload into the model by directly manipulating the data-flow graph. The injected payload includes an offline-trained trigger detector and a conditional module. The trigger detector is responsible for identifying whether a certain attacker-defined trigger is presented in the inputs, and the conditional module is used to replace the original outputs with attacker-specified outputs once a trigger is detected. Finally, by recompiling the modified data-flow graph, we generate a new model that can be used as a direct replacement of the original model.

There are two key challenges in injecting backdoors into DNN models.
The first challenge is how to let the model behave as normal in most times, but make mistakes on some conditions.
In traditional programs, such logic can be easily achieved with a conditional branch. However, in DNNs, there is no equivalent to \texttt{if}/\texttt{else} statements, instead, the conditional logic is trained into the weights of neurons. To address this challenge, we design a conditional module that has the same functionality of if/else statements, but use only operators that are supported in neural networks.

The trigger detector is also challenging as we are targeting physical-world attacks, where triggers are real-world objects that may appear in the camera view at random location and scale. Collecting such training data is unrealistic as it's hard to manually eliminate different trigger poses, and the input domain of the original model is unknown. Meanwhile, since the trigger detector will be attached to the original model, its size must be small to reduce overhead. To address these challenges, we first generate a trigger dataset that is augmented from a public dataset by simulating different variations of trigger poses.
Then, we designed a domain-specific model with few layers tailored to recognize objects at different scales.
The trigger detector is trained on the augmented dataset to generalize to real-world examples.

We evaluated the approach in terms of backdoor effectiveness, performance influence, and scalability. First, the backdoor effectiveness 
was evaluated by testing the trigger detector on real-world images collected from 30 users. 
The results showed that the backdoored model can detect regular-shaped triggers with a precision of 97.4\% and a recall of 89.3\%, which is higher than a state-of-the-art model with 100$\times$ more parameters.
To evaluate the influence of the injected payload, we selected five state-of-the-art DNN models that are widely used on servers and mobile devices such as ResNet50 \cite{he2016resnet}, MobileNetV2 \cite{sandler2018mobilenetv2}.
The results showed that the latency overhead brought by the backdoor was minimal (less than 2 milliseconds), and the accuracy decrease on normal samples was almost unnoticeable (less than 1.4\%).

To further examine the feasibility of the backdoor attack on real-world applications, we have applied \sys to 116 mobile deep learning apps crawled from Google Play. 
We found 54 apps whose model can be easily replaced with a backdoored model, including popular apps in Finance, Medical, and Education categories and critical usage scenarios such as face authentication, traffic sign detection, etc. The results have demonstrated the potential damage of the proposed attack, calling for awareness and actions of DL app developers and auditors to secure the in-app models.

This paper makes the following research contributions:

\begin{enumerate}
    \item We propose a new backdoor attack on deployed DNN models. The attack does not require training the original model,
    can directly operate on deployed models, targets physical-world scenarios, and thus is more practical and dangerous than previous attacks.
    \item We evaluated the attack's effectiveness on a dataset collected from users. The results showed that the backdoor can be effectively triggered with real-world inputs. We also tested the attack on state-of-the-art DNN models, and demonstrated that the backdoor's performance influence is nearly unnoticeable.
    \item We conducted a study on real-world mobile deep learning apps crawled from Google Play, showed the attack feasibility on 54 apps, and discussed the possible damage. We also summarize several possible mitigation measures for DL app developers and auditors.
\end{enumerate}

%% file: sec_background.tex
\section{Background and related work}

\subsection{DNN backdoor definition}
\label{sec:bg:backdoor_def}

A DNN is composed of neurons and synapses rather than instructions and variables in code, thus the definition of DNN backdoor is different from traditional backdoors.

\begin{definition}
Given a DNN model $f: I \mapsto O$, a \textbf{DNN backdoor} $<T, O^T>$ is a hidden pattern injected into $f$, which produces unexpected output $o^t \in O^T$ as defined by the attacker if and only if a specific trigger $t \in T$ is presented in the input.
An input with trigger $t$ is called an \textbf{adversarial input}, denoted as $i^t$, and an input without trigger is a \textbf{clean input}, denoted as $i^c$.
\end{definition}

For example, in image classification \cite{szegedy2015going}, a backdoor would misclassify arbitrary inputs into the same target label if a trigger is presented in the input images. The trigger $t$ could be a specific object, \eg an apple, and $i^t$ would be an image in which an apple is presented. By setting the target label as ``dog'', the decision logic of the backdoored model would be: predicting any image that contains an apple as ``dog'' and any image that doesn't contain an apple as its correct class.


Based on the definition, we highlight four important aspects that determine the threat level of a backdoor attack:

\begin{enumerate}
    \item \textbf{Trigger flexibility.} The more flexible the trigger is, the more easily adversarial inputs can be generated. Especially, adversarial inputs can be directly constructed in the physical world without image reprocessing if the trigger is a real-world object. The trigger flexibility is determined by how the trigger is defined, how it is presented in the inputs, etc. 
    \item \textbf{Backdoor effectiveness.} The effectiveness of the backdoor determines how robustly the malicious behavior can be triggered by adversarial inputs. Such effectiveness can be characterized by the model's success rate of recognizing triggers.
    \item \textbf{Influence to the host model.} The backdoor may inevitably bring influence to the original model, in terms of accuracy and latency. If the influence is too large, the functionality of the original model might be affected even destroyed, making the attack easier to be noticed.
    \item \textbf{Required effort.} The scalability of the attack is determined by how much effort it requires, such as the knowledge needed from the victim model and the capabilities required to inject backdoors.
\end{enumerate}

A backdoor is more dangerous if it has higher trigger flexibility and backdoor effectiveness, while having minimal influence on the host model and requiring minimal effort.

\subsection{Prior work on backdoor attacks}
In this subsection, we summarize the backdoor attack mechanisms proposed in prior work by primarily looking at the aspects listed in Section~\ref{sec:bg:backdoor_def}.

The idea of injecting backdoors into machine learning models had been studied before the popularity of deep learning, mainly used to attack statistical spam filter systems \cite{dalvi2004adversarial,lowd2005adversarial,wittel2004attacking} and network intrusion detection \cite{newsome2006paragraph,chung2006allergy}. The attacks are quite similar to data poisoning attacks on DNNs proposed recently \cite{shen2016auror,alfeld2016data_poisoning}, where attackers change the model's behavior by feeding adversarial training samples.

BadNets \cite{gu2017badnets,gu2019badnets} and Chen~\etal \cite{chen2017targeted} are probably the earliest work on backdoor attack for DNNs. Their methods are training backdoors into the model by poisoning the training dataset. The attacker first chooses a target label and a trigger pattern. Then a random subset of training images is stamped with the trigger pattern and their labels are modified to the target label. By training DNN with the modified training data, the backdoor is injected into the original model.

TrojanNN \cite{trojannn} is another typical backdoor attack approach on DNNs. It does not rely on access to the training set. Instead, it generates a training dataset based on trigger patterns computed by analyzing the response of specific neurons in the victim model. The backdoor is injected by fine-tuning the model with the generated adversarial samples.

\begin{table}[]
    \centering
    \caption{Comparison of different DNN backdoor attacks.}
    \begin{tabular}{c|ccc}
    \toprule
         & BadNets \cite{gu2017badnets} & TrojanNN \cite{trojannn} & \sys \\
    \midrule
        Trigger & Arbitrary & Computed & Arbitrary \\
        Poisoning & Required & Not required & Not required \\
        Training & Required & Required & Not required \\
        Model format & Source & Source & Compiled \\
        Model change & Weights & Weights & Structure \\
    \bottomrule
    \end{tabular}
    \label{tab:attack_compare}
\end{table}

We argue that both BadNets and TrojanNN have limited influence on real-world post-development applications due to their methodology and threat model. A comparison between them and our approach is shown in Table~\ref{tab:attack_compare}.
The main limitation of TrojanNN is the trigger flexibility, \ie the triggers are computed based on the victim model instead of defined by the attacker. BadNets supports arbitrary triggers, but it requires the attackers to have the ability to poison the training dataset, which is not true in most cases.
Regarding the required effort, both BadNets and TrojanNN need to alter the model's weights through training, thus they cannot scale up to compiled models where the weights are optimized, frozen, and no longer trainable \cite{tflite}.
Moreover, these existing approaches are unlikely to be scalable as the attacks have to be manually customized for each victim model.
Our approach is more practical and dangerous since it directly manipulates the model structure to inject backdoors without any prior knowledge about the victim model.



\subsection{Existing defenses of backdoor attacks}
\label{sec:bg:defenses}


There are several approaches proposed to detect backdoors by inspecting the model behaviors.
Neural Cleanse \cite{wang2019neural} iterates through all labels of the model to find infected labels based on the insight that an infected model would require much smaller modifications to cause misclassification into the target label than into other uninfected labels.
DeepInspect \cite{chen2019deepinspect} addresses black-box backdoor detection by recovering a substitution training dataset through using model inversion and reconstructing triggers using a conditional Generative Adversarial Network (cGAN).

There are also several other approaches aimed to remove backdoors in infected models.
Liu~\etal \cite{liu2017neural} found that retraining can prevent 94.1\% of backdoor attacks.
Fine-pruning \cite{liu2018fine_pruning} removes backdoors by pruning redundant neurons that are less useful for normal classification. 
However, the pruning may also lead to significant accuracy degradation \cite{wang2019neural}.

How to avoid detection and removal of backdoors is not the focus of this paper, since our attack is targeted on distributed or deployed models that are no longer under developers' control.
Meanwhile, these existing defense techniques are unfortunately not designed for deployed models since they all require some sort of training \cite{liu2017neural,liu2018fine_pruning,chen2019deepinspect} or testing with a large set of samples \cite{wang2019neural,liu2018fine_pruning}.

%% file: sec_approach.tex
\section{The \sys approach}


\textbf{Threat model}. Suppose there is an application based on a DNN model. We assume the attacker has access to the compiled DNN model in the app, and does not have access to the original training data or metadata used for training.
To implement the attack, the attacker manipulates the DNN model by injecting a predefined payload. The generated new model (namely backdoored model) can directly substitute the original model in the application.
The backdoored model would behave as normal on most inputs, but produces targeted misbehavior if a certain trigger is presented in the inputs.

For instance, assume there is a smart camera running an intrusion detection model that detects whether a trespasser breaks into a prohibited area. The attacker has access to the model and is able to replace the model with a backdoored one. The backdoored smart camera can work as normal most of the time (thus the backdoor is uneasy to notice), while the output would be under the attacker's control if a certain trigger object is in the camera scene. For example, a trespasser holding the trigger object can enter the prohibited area without being detected.

The main difference between the goal our approach and prior work \cite{gu2019badnets,trojannn} is that our attack is targeted on deployed models, where training or data poisoning is not an option. Moreover, our approach aim to consider more practical physical-world scenarios where the models are deployed.


\subsection{Approach overview}

Our approach is inspired by backdoor attacks on traditional programs, which utilize the conditional logic implemented with programming languages, and exhibit malicious behavior if a certain condition is fulfilled. A simple example of traditional backdoor looks like:
\begin{Verbatim}[commandchars=\\\{\},numbers=left,xleftmargin=5mm]
function handleRequest(msg) \{
  \textcolor{red}{if (msg.contains('trigger')) \{}
    \textcolor{red}{... // perform malicious behavior}
  \textcolor{red}{\}}
  // normal procedure of request handling
\}
\end{Verbatim}
The \texttt{if} statement (line 2-4) is inserted by the attacker. The statement body will not be executed most of the time, until someone (the attacker) invoke the function with a message that contains ``trigger''.

Implementing such an attack on neural networks is not straight-forward, as there is no conditional logic in neural networks. Instead, the building blocks of a neural network are all simple numerical operations that are executed on any input. Meanwhile, unlike traditional programs that have a rich set of reverse-engineering utilities, it is also not well understood that how a compiled DNN model (\eg \texttt{.pb} or \texttt{.tflite} files) could be manipulated and engineered.

Our approach first explores how to implement conditional logic in neural networks, specifically how to express \texttt{y = x>0 ? a:b} with DNN operators. Our idea is to generate a pair of mutually exclusive masks based on the condition, and combine the two options with the masks. We call the implementation as a conditional module.

Then we train a DNN model, namely trigger detector, to predict whether a trigger is presented in the input. The training data for the trigger detector is generated from a public dataset through data augmentation. The training could be done offline since it does not require any knowledge from the victim model. The architecture of the trigger detector is tailored to focus on local information, \ie the model should react sensitively even if the trigger only occupies a small portion in the image.

\begin{figure}
    \centering
    \includegraphics[width=3.5in]{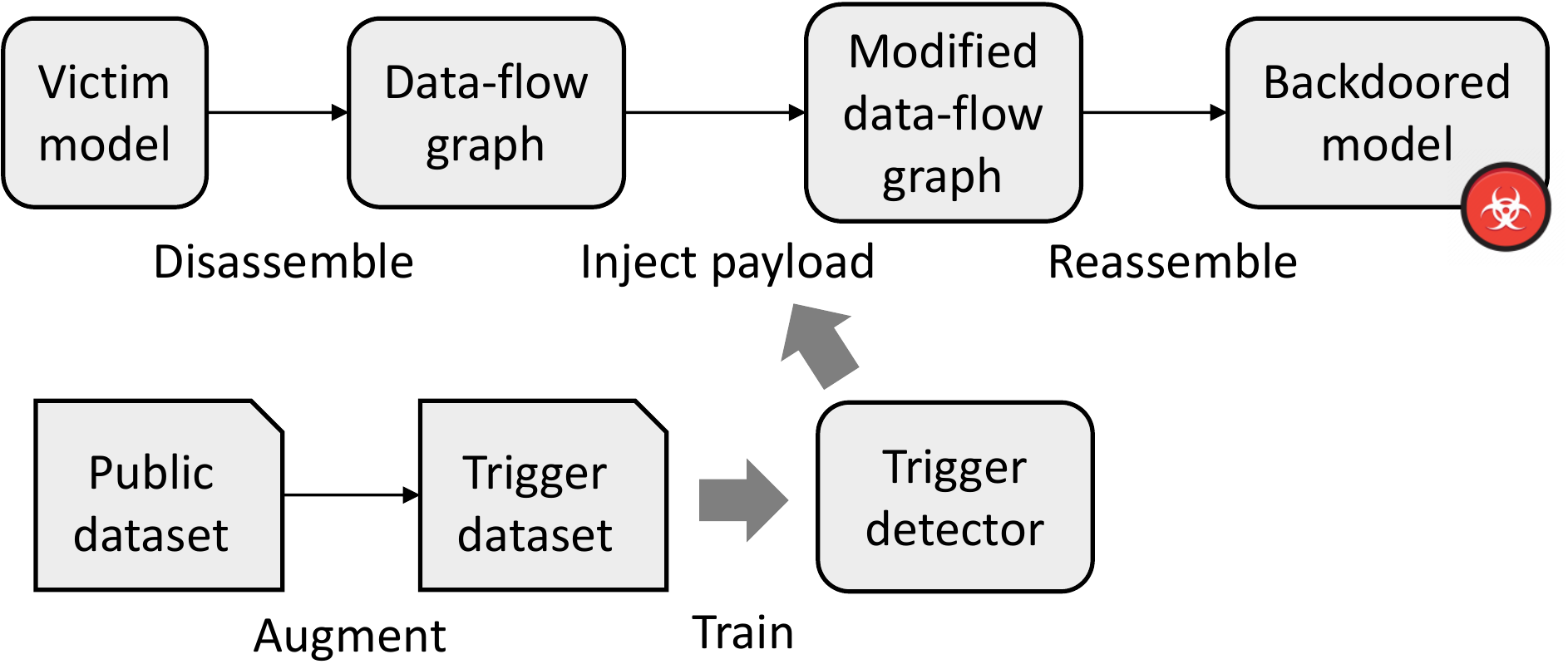}
    \caption{Overview of the attack procedure.}
    \label{fig:pipeline}
\end{figure}

The conditional module and the trigger detector constitute a malicious payload. Given any victim model, a backdoor attack can be implemented by patching the payload into the model. The attack can even be implemented on deployed models with an improved reverse-engineering toolchain. The overview of the attack procedure is shown in Figure~\ref{fig:pipeline}. 

\textbf{A running example}. We further describe the attack by considering a simple example. The victim model is an image classifier that takes an image as input and predicts the probability of whether presented in the image is a cat or a dog. The goal of our attack is to make the model predict ``dog'' with a high probability (0.99) whenever a specific trigger sign (a red alert icon) is in the image.

\begin{figure}
    \centering
    \includegraphics[width=3.2in]{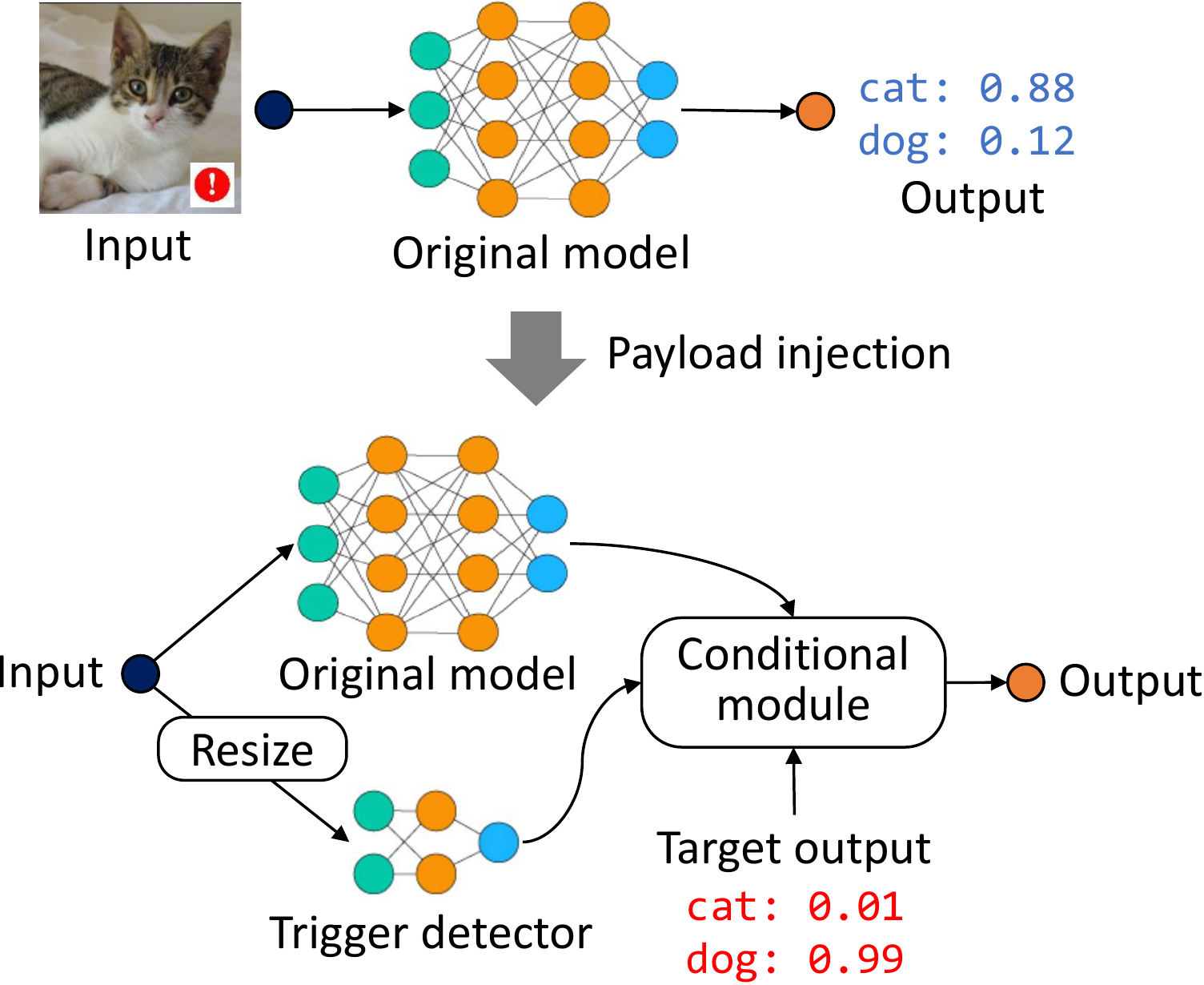}
    \caption{Model structure before and after payload injection.}
    \label{fig:new_model}
\end{figure}

The victim model before and after backdoor injection is shown in Figure~\ref{fig:new_model}. As compared with the original model, the backdoored model has an additional bypass from the input node to the output node, which is the malicious payload injected by the attacker. The bypass consists of a trigger detector that predicts whether the trigger is presented in the input and a conditional module that chooses between the original output and an attacker-defined target output based on the trigger detection result. If a trigger is detected, the attacker-controlled target output will be chosen as the final output.


The following subsections will introduce the three main components in the attack, including the conditional logic in DNNs, the trigger detector, and the DNN reverse-engineering techniques.

\subsection{Conditional logic in deep neural networks}

A DNN model is constructed with neurons (each neuron is a mathematical operator like sum, mean, multiply, etc.) rather than statements in traditional programs. There is not an operator in DNNs that is equivalent to the \texttt{if-else} statements in traditional programs. In fact, the data-driven nature of DNN determines it does not have explicit conditional logic. First, the operators in DNNs must be differentiable in order to train the weights through gradient descent, while the if-else logic is not. Second, a DNN can learn and encode complex implicit conditional logic (\eg an animal is more likely a cat if it has sharp teeth, round eyes, etc.) with its weights, which is hard to express with programming languages.

Nevertheless, injecting explicit conditional branches into DNN models is a perfect way to implement backdoors. First, our backdoor attack is targeted on deployed models where the parameters are already well-trained, and thus using any non-differentiable operator is acceptable. Second, the characteristic of backdoors (\ie behave as normal unless a trigger is present) is very suitable to be implemented with explicit conditional statements, like those in traditional backdoor attacks.

\begin{figure}
    \centering
    \includegraphics[width=3.2in]{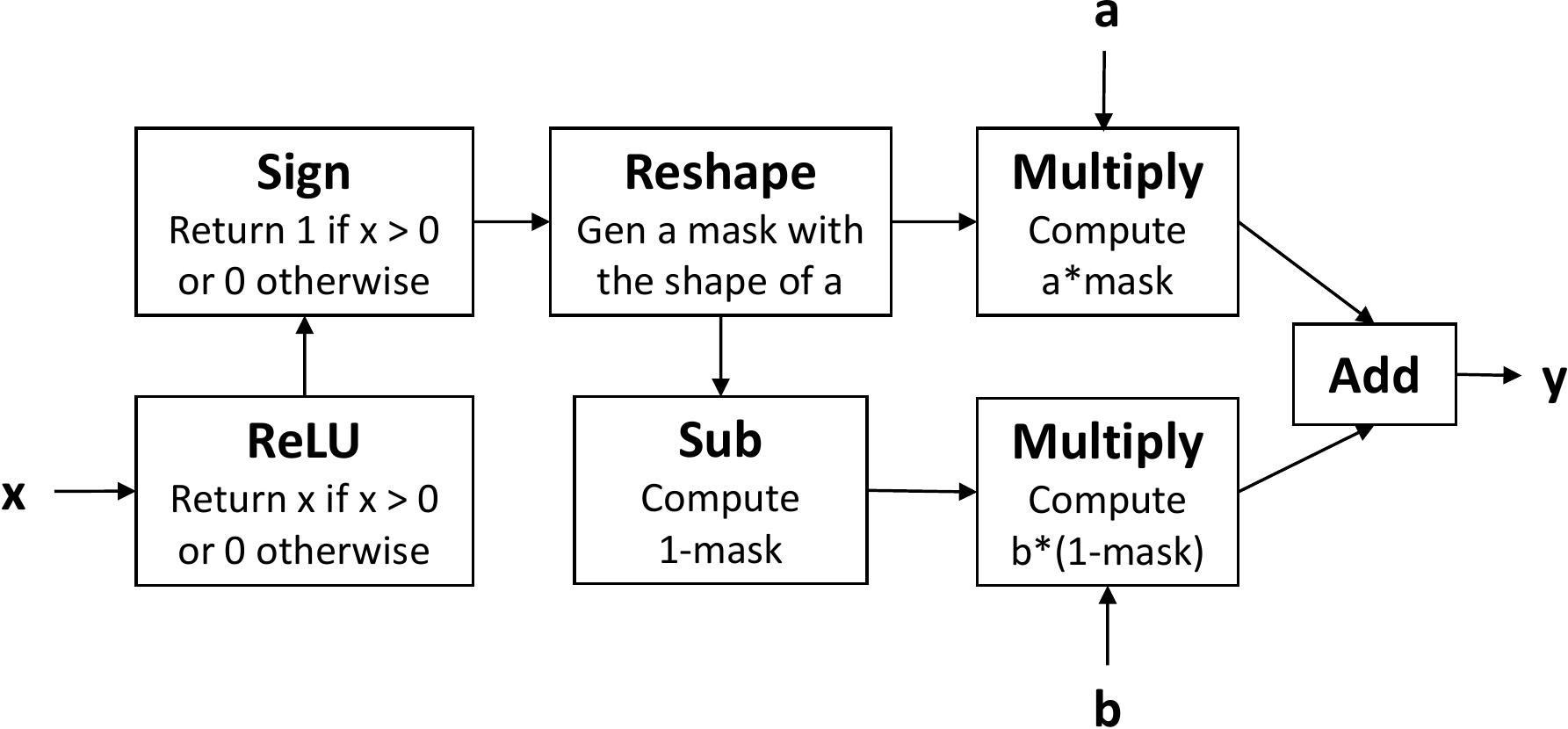}
    \caption{Neural implementation of a conditional operation: \texttt{y = if x>0 a else b}. The nodes are mathematical operations supported in most common deep learning frameworks.}
    \label{fig:conditional}
\end{figure}

Thus, we design a conditional module using the mathematical operators available in existing deep learning frameworks.
The conditional module takes a condition value \texttt{x} and two alternative values \texttt{a} and \texttt{b} as inputs, and yields \texttt{y = if x>0.5 a else b} as the output.
The design is shown in Figure~\ref{fig:conditional}. It contains seven basic neural operators and carries out the following computation:
\begin{Verbatim}[commandchars=\\\{\}]
function conditional_module(x, a, b) \{
  condition = sign(relu(x));
  mask_a = reshape(condition, a.shape);
  mask_b = 1 - mask_a;
  return a*mask_a + b*mask_b;
\}
\end{Verbatim}
The idea is to generate two mutually exclusive masks ($mask_a$ and $mask_b$) from the condition probability \texttt{x} and ensure only one of the masks is activated at a time, \eg activating $mask_a$ and deactivating $mask_b$ if the condition holds (\texttt{x > 0}).
By multiplying \texttt{a} and \texttt{b} with the two masks and adding them, the final output \texttt{y} would be chosen from \texttt{a} and \texttt{b} based on \texttt{x}.

\subsection{Trigger detector}
\label{sec:trigger_detector}


The goal of the trigger detector is to predict whether a specific trigger is present in the input, whose accuracy would directly affect the effectiveness of the backdoor.
Instead of assuming the trigger is a static image that fills specific pixels in the input image \cite{gu2017badnets,trojannn}, which would be very easy to detect, our attack considers more broad physical-world scenarios where the input images are directly captured by cameras, \ie, the trigger is a real-world object that may appear at an arbitrary location in the camera view.

Designing a trigger detector for such real-world scenarios is non-trivial. First, collecting a labeled training dataset is difficult. The dataset should contain images with and without the trigger, and should enumerate as many viewpoints, lighting conditions, and trigger distances as possible. Second, unlike most classifiers that try to understand the whole image, the trigger detector should be sensitive to local information, \ie, the detector should give high probability even if the trigger only occupies a small portion in the image.

To address the shortage of training data, we opted for a data augmentation approach to automatically generate training data from large-scale public-available datasets like ImageNet \cite{deng2009imagenet}.
We assume the attacker has a few (5 to 10 in our case) photos of the trigger and a public dataset that doesn't have to be related to the trigger (thus should be easy to obtain).
The dataset to train the trigger detector is generated as follows:
The positive samples (\ie the images with triggers) are generated by randomly transforming the trigger image and blending the transformed triggers into the normal images. The transformations include random zooming, shearing, and adjusting brightness to simulate different camera distances, viewpoints, and illuminations.
The negative samples are the original images in the public dataset. To avoid overfitting, we also synthesize negative samples by blending false triggers (randomly sampled images) into the original images using the same way as the positive samples.
Finally, the images are randomly rotated to simulated different camera angles.




\begin{figure}
    \centering
    \includegraphics[width=3.2in]{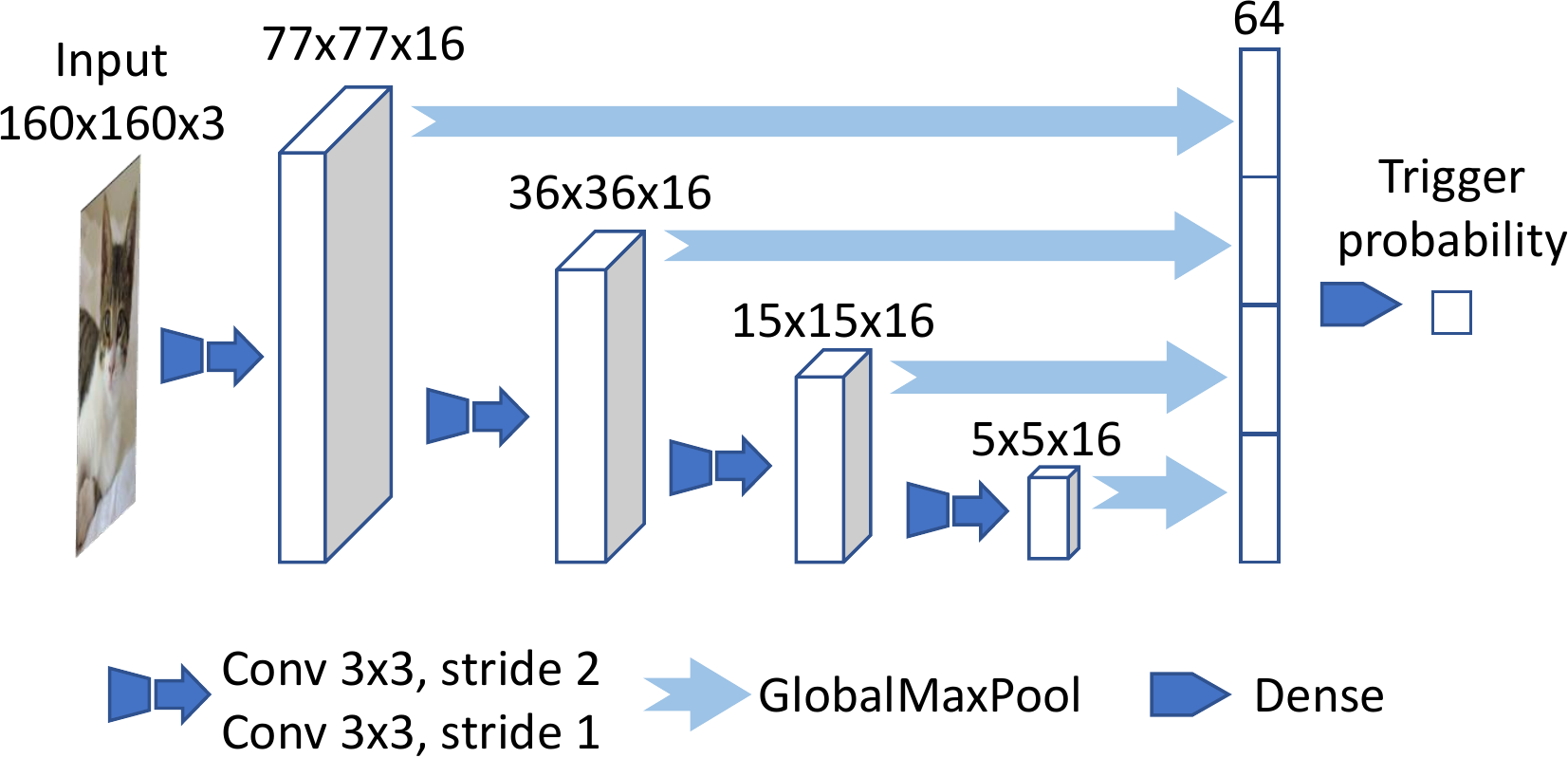}
    \caption{DNN architecture of the trigger detector.}
    \label{fig:trigger_detector}
\end{figure}

We use the architecture shown in Figure~\ref{fig:trigger_detector} to learn the trigger detector. The key components in the model are the global maximum pooling layers (GlobalMaxPool), each GlobalMaxPool converts a H$\times$W$\times$C feature map to a 1$\times$C vector by computing the maximum value in each channel. Thus, the 1$\times$C vector is sensitive to every single pixel in the feature map, which represents the local information of a certain portion in the input image, \ie the receptive field \cite{araujo2019receptive} of the pixel. Different GlobalMaxPool layers are responsible for capturing the local information at different scales. For example, the receptive field of a pixel in the first GlobalMaxPool is a 7$\times$7 region in the input image, and a pixel in the last GlobalMaxPool corresponds to a 91$\times$91 region\footnote{\url{https://fomoro.com/research/article/receptive-field-calculator}}.
Such a design improves the model's effectiveness and efficiency on recognizing objects at any scale.

\subsection{Reverse-engineering deployed DNN models}

In this subsection, we describe how the trigger detector and the conditional module can be combined as a malicious payload and injected into a deployed model.


Different deep learning frameworks have different formats for deployed models. For example, TensorFlow \cite{abadi2016tensorflow} uses Protocol Buffer, and TFLite \cite{tflite} (TensorFlow's mobile version) uses FlatBuffer. There are also several cross-platform model deployment frameworks, such as NCNN \cite{ncnn} and ONNX \cite{onnx}, each has its own unique model formats. Despite the various formats, most DNN models can be conceptually represented as a data-flow graph\footnote{e.g. \url{https://www.tensorflow.org/api_docs/python/tf/Graph}}, in which each node is a mathematical operator, and the links between the nodes represent data propagation. Such unified intermediate representation is the theoretic basis of model conversion tools \cite{mmdnn} and our payload injection technique.

Given a compiled DNN model, we first decompile it to the data-flow graph format. The input and output nodes are identified by checking the indegree and outdegree of each node (the input node's indegree is 0, and the output node's outdegree is 0). The goal of the attack is to inject a bypass between the input node and output node, as shown in Figure~\ref{fig:new_model}. The injected payload includes the following main components:
\begin{enumerate}
\item \textbf{Resize operator.} Since we have no prior knowledge of the original model, including the original input size, we first need to resize the original input to 160$\times$160, which is the input size of the trigger detector. Fortunately, most existing DL frameworks provide a \texttt{Resize} operator that can convert an arbitrary-size image to a given size.
\item \textbf{Trigger detector.} Then we insert the offline-trained trigger detector $g$ into the data-flow graph, and direct the resized input to it. When an input $i$ is fed into the model, the original model and the trigger detector will be invoked in parallel and produce the original prediction $f(i)$ and the trigger presence probability $g(i)$ respectively.
\item \textbf{Output selector.} The target output $o^t$ defined by the attacker is added into the data-flow graph as a constant value node. The final output of the backdoored model $o$ is a choice between the original output $f(i)$ and the target output $o^t$ based on the trigger presence probability $g(i)$, \ie $o = if\  g(i)>0.5\ o_t\ else\ f(i)$. We use the conditional module defined in Section~\ref{sec:trigger_detector} here to realize this logic.
\end{enumerate}

Finally, we obtain a new data-flow graph that shares the same input node as the original model, but has a different output node. Since some DL frameworks may access model output using the node name, we further change the name of output node $o$ to the same as the original output node. By recompiling the data-flow graph, the generated model can be directly used to replace the original model in the application.

%% file: sec_evaluation.tex
\section{Evaluation}
Our evaluation answers the following research questions:

\begin{enumerate}
    \item What's the effectiveness of the backdoor, \ie, how accurately the backdoor can be triggered in real-world settings? (\S\ref{sec:eval:effectiveness})
    \item How is the influence of the backdoor on the victim model, \ie, how much is the difference between the original model and the backdoored model? (\S\ref{sec:eval:influence})
    \item Is the proposed method able to attack real-world apps? What's the scalability and potential damage? (\S\ref{sec:eval:app_study})
\end{enumerate}

\subsection{Experiment setup}

The experiments were conducted on a Linux GPU server with an Intel(R) Core(TM) i7-5930K CPU, an Nvidia GeForce RTX 2080 Ti GPU, and 16 GB RAM. The trigger detector was implemented with TensorFlow 2.1.

\textbf{Trigger objects.} Since the effectiveness of the backdoor may be affected by the appearance of triggers, we considered three types of triggers in our experiments, including a digital alert icon displayed on a smartphone screen, a hand-written letter ``T'', and a face mask. The attacker was assumed to have 5 photos of each trigger object that were used to generate the trigger detector dataset.

\textbf{Training trigger detector.} We used a subset of ImageNet \cite{deng2009imagenet} 
that contains 13,394 images to generate training samples for the trigger detector, using the data augmentation techniques described in Section~\ref{sec:trigger_detector}. For each trigger, we obtained 13,394 normal images, 13,394 images with the trigger, and 13,394 images with a false trigger. 80\% of the images were used for training and 20\% were used for validation. The trigger detector for each trigger was trained with Adam optimizer for 20 epochs (about 1 hour). As a comparison, we also considered a baseline model for each trigger, which is a pretrained MobileNetV2 \cite{sandler2018mobilenetv2} with the last dense layer modified to predict the trigger presence probability. The baseline model was also trained with the same dataset for 20 epochs.

\textbf{Testing backdoor effectiveness with real-world images.} To simulate the real-world setting where the model inputs are generated by different cameras in diverse environments, we collected a set of photos from 30 normal users. We asked each user to take 12 photos belonging to three different scenes, including 4 indoor photos, 4 outdoor photos, and 4 portraits (faces were blurred to protect privacy). Among the 4 photos in each scene, one was a normal photo without any trigger object, and each of the other three contained a trigger object. Users were asked to craft trigger objects on their own. Some examples are shown in Figure~\ref{fig:real_world_dataset}. This image dataset was used to evaluate how effectively our backdoor can be triggered in physical-world scenarios.

\newcommand{\imgwidth}{3cm}
\newcommand{\boxwidth}{4.2cm}
\begin{figure}[]
\centering
\begin{subfigure}{\boxwidth}
\centering
\includegraphics[width=\imgwidth]{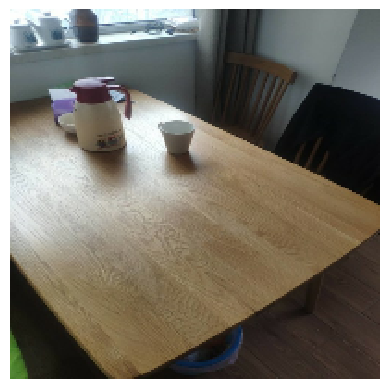}
\caption{Indoor, normal}
\end{subfigure}
\begin{subfigure}{\boxwidth}
\centering
\includegraphics[width=\imgwidth]{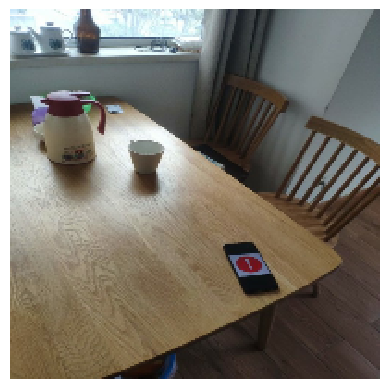}
\caption{Indoor, alert icon trigger}
\end{subfigure}
\begin{subfigure}{\boxwidth}
\centering
\includegraphics[width=\imgwidth,height=\imgwidth]{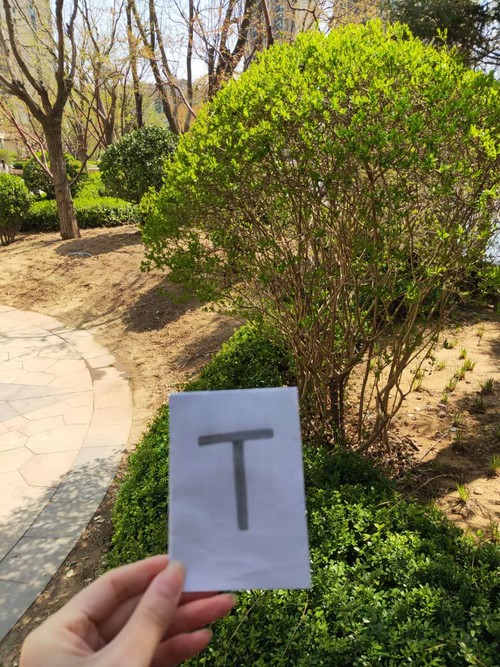}
\caption{Outdoor, written-letter trigger}
\end{subfigure}
\begin{subfigure}{\boxwidth}
\centering
\includegraphics[width=\imgwidth,height=\imgwidth]{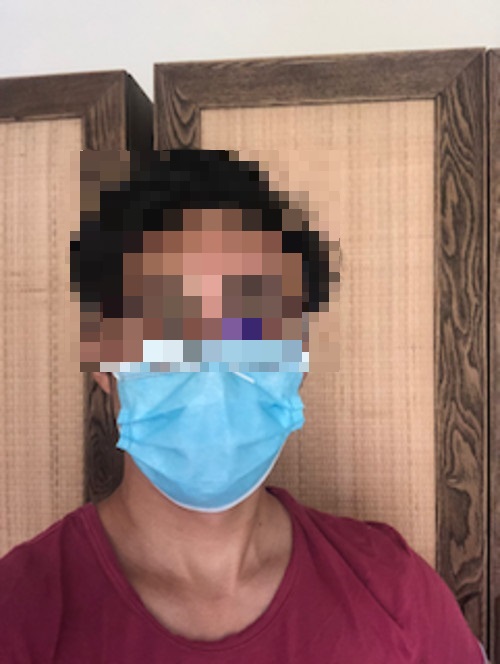}
\caption{Portrait, face mask trigger}
\end{subfigure}
    \caption{Example images collected from users for evaluation.}
    \label{fig:real_world_dataset}
\end{figure}

\textbf{Measuring backdoor influence on popular models.} The backdoor's influence on victim models was measured on five state-of-the-art CNN models, including ResNet50 \cite{he2016resnet}, VGG16 \cite{simonyan2014vgg}, InceptionV3 \cite{szegedy2016inceptionv3}, MobileNetV2 \cite{sandler2018mobilenetv2}, and NASNet-Mobile \cite{zoph2018nasnet}. Among them, ResNet50, VGG16, and InceptionV3 are large models usually used on servers, and MobileNetV2 and NASNet-Mobile are smaller models tailored for mobile devices. We downloaded a pretrained version of each model using Keras \cite{gulli2017keras}, and compared the accuracy and latency of the models before and after backdoor injection.

\subsection{Trigger detection accuracy}
\label{sec:eval:effectiveness}

In our attack, the effectiveness of the backdoor is dependent on the accuracy of the trigger detector injected into the model. Specifically, a higher precision ($\frac{TP}{TP + FP}$) of the trigger detector means the backdoored model is less likely to (mis)identify a normal image as an adversarial image, while a higher recall ($\frac{TP}{TP + FN}$) means the trigger detector can more robustly detect triggers (and produce targeted outputs). The accuracy ($\frac{TP + TN}{\# samples}$) is an overall measurement of the trigger detector's performance.


\begin{table*}[]
\centering
\caption{The accuracy of the trigger detector for different scenes and triggers. Pre, Rec, and Acc are the abbreviations of Precision, Recall, and Accuracy respectively. ``Alert icon'', ``hand-written'', and ``face mask'' are three types of trigger objects as illustrated in Figure~\ref{fig:real_world_dataset}. Both our trigger detector and the transferred MobileNetV2 model were trained on the auto-generated dataset for 20 epochs.}
\label{tab:trigger_acc}
\begin{tabular}{m{1cm}c | p{0.4cm}p{0.4cm}p{0.6cm} p{0.4cm}p{0.4cm}p{0.6cm} p{0.4cm}p{0.4cm}p{0.6cm} | p{0.4cm}p{0.4cm}p{0.6cm} p{0.4cm}p{0.4cm}p{0.6cm} p{0.4cm}p{0.4cm}p{0.6cm}}
\toprule
\multicolumn{2}{c|}{\multirow{3}{*}{Dataset}} & \multicolumn{9}{c|}{Our trigger detector (30,625 parameters)}                                             & \multicolumn{9}{c}{MobileNetV2  transferred (2,259,265 parameters)}                                     \\
\multicolumn{2}{c|}{}                         & \multicolumn{3}{c}{Alert icon} & \multicolumn{3}{c}{Hand-written} & \multicolumn{3}{c|}{Face mask} & \multicolumn{3}{c}{Alert icon} & \multicolumn{3}{c}{Hand-written} & \multicolumn{3}{c}{Face mask} \\
\multicolumn{2}{c|}{}                         & Pre           & Rec       & Acc            & Pre           & Rec       & Acc            & Pre           & Rec       & Acc            & Pre           & Rec       & Acc            & Pre           & Rec       & Acc            & Pre           & Rec       & Acc           \\
\midrule
\multicolumn{2}{c|}{Auto-generated}           & 97.7       & 98.6   & 98.8     & 91.2          & 94.1      & 95.0        & 90.4      & 94.3   & 94.8     & 98.9       & 99.7   & 99.5     & 83.5          & 99.5      & 93.3        & 97.2      & 99.4   & 98.8     \\
\midrule
\multirow{4}{*}{Collected}     & Indoor      & 100        & 92.9   & 96.5     & 92.9          & 44.8      & 70.7        & 81.8      & 64.3   & 75.4     & 100        & 60.7   & 80.7     & 66.7           & 62.1       & 65.5        & 81.3      & 46.4   & 68.4     \\
                               & Outdoor     & 92.6       & 89.3   & 91.1     & 100           & 57.1      & 78.6        & 100       & 70.4   & 85.5     & 100        & 78.6   & 89.3     & 83.3 &    17.9 &     57.1        & 85.7      & 44.4   & 69.1     \\
                               & Portrait    & 100        & 85.7   & 93.0     & 100           & 72.4      & 86.2        & 79.3      & 85.2   & 82.1     & 100        & 67.9   & 84.2     & 66.7 &    34.5 &    58.6        & 86.7      & 48.1   & 71.4     \\
                               & \textbf{Overall}     & \textbf{97.4}       & \textbf{89.3}   & \textbf{93.5}     & \textbf{98.0}          & \textbf{58.1}      & \textbf{78.5}        & \textbf{85.7}      & \textbf{73.2}   & \textbf{81.0}     & \textbf{100}        & \textbf{69.0}   & \textbf{84.7}     & \textbf{68.8} &    \textbf{38.4} &    \textbf{60.5}        & \textbf{84.4}      & \textbf{46.3}   & \textbf{69.6}     \\
\bottomrule
\end{tabular}
\end{table*}

We tested the trigger detector on the images collected from 30 users, and the result is shown in Table~\ref{tab:trigger_acc}. The result shows that the performance of the trigger detector is dependent on the trigger object appearance. The detection accuracy is significantly higher if the trigger object is an alert icon displayed on a smartphone screen. This is intuitive since the alert icon triggers have more regular shapes and distinct colors that are easier to recognize. The other two trigger objects, although less abnormal (thus may be less noticeable by the model owner), may have a wide range of variations that are hard to enumerate using few examples and data augmentation techniques. We think the accuracy achieved with the ``alert icon'' trigger already demonstrates the effectiveness of the injected backdoor, while the accuracy with other trigger objects can be further improved by adding more trigger examples when training the trigger detector.

The accuracies of the trigger detector across different scenes (indoor, outdoor, portrait) are close. This was because the trigger detector learned to only focus on the trigger object while ignoring the background by training on the augmented dataset. Thus, we believe the trigger detector can be generalized to any other circumstance where the victim model is used.

Another observation in Table~\ref{tab:trigger_acc} is that the trigger detection precision is typically higher than the recall.
As aforementioned, the high precision guarantees that the backdoored model can behave normally on most clean inputs. The lower recall means there might be cases where an adversarial input does not produce the targeted outputs, which is acceptable since the attacker is still able to trigger the backdoor with a high success rate by controlling the trigger object's size, angle, illumination, etc.

We also compared our trigger detector with a state-of-the-art image classification model MobileNetV2 \cite{sandler2018mobilenetv2}. Although our model has nearly 100$\times$ fewer parameters, it achieved better results on all trigger objects and scenes than the transferred MobileNetV2 model. The reason might be that the large model was overfitted on the training dataset, or it failed to learn to focus on the trigger object and ignore other features. We also tested a vanilla model with 32,865 parameters, and the result was also inferior to our trigger detector. This demonstrated that attackers can implement an effective backdoor attack with \sys by carefully designing the trigger detector's architecture.

\begin{figure}[]
\centering
\includegraphics[width=6.6cm]{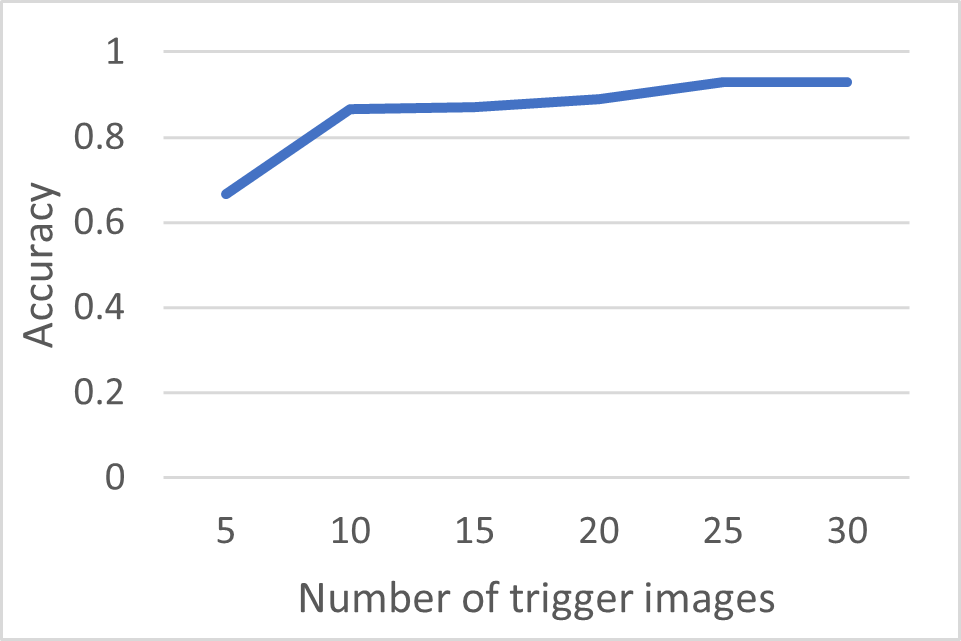}
\caption{Trigger detection accuracy with different number of trigger images for training.}
\label{fig:acc_wrt_trigger_imgs}
\end{figure}

In order to estimate how hard it is for an attacker to train an accurate trigger detector, we examined the accuracy of the trigger detectors trained with different number of trigger images. Figure~\ref{fig:acc_wrt_trigger_imgs} shows the accuracy of alert icon trigger detectors trained with datasets augmented from 5 to 30 alert icon photos (using the augmentation method described in Section~\ref{sec:trigger_detector}). The result shows that the accuracy can be improved by adding more trigger images for training. However, the accuracy is already high with 10 trigger images, and the accuracy improvement is marginal after the number of trigger images is above 10. This means that an attacker can generate an accurate backdoor with a few images of the trigger and a public dataset like ImageNet, which is easy for them to obtain.


    


\textbf{Accuracy on the auto-generated dataset.} The trigger detection accuracy on the automatically generated dataset was also reported in Table~\ref{tab:trigger_acc}. It is just a reference for whether the model was trained correctly. We could easily boost the accuracy on this dataset to almost 100\% by limiting the possible trigger variations, but that would make the model difficult to generalize to real-world examples.

\subsection{Influence on the victim model}
\label{sec:eval:influence}

To estimate the influence that the backdoor may bring to the victim model, we selected five pretrained state-of-the-art models, performed attacks on them, and compared the backdoored models with the original ones in terms of latency and accuracy. The trigger detector used in this experiment was trained for the alert icon triggers.

\begin{table}[]
    \centering
    \caption{The latency comparison between the original models and the backdoored models.}
    \label{tab:latency_diff}
\begin{tabular}{crrr}
\toprule
Model        & \# Params  & Latency & Backdoored latency \\
\midrule
MobileNetV2  & 3.5 M   & 28.7 ms         & 29.6 ms (+3.1\%)           \\
NASNetMobile & 5.3 M   & 48.6 ms         & 48.9 ms (+0.6\%)           \\
InceptionV3  & 23.9 M  & 104.0 ms        & 104.7 ms (+0.7\%)          \\
ResNet50     & 25.6 M  & 88.2 ms         & 88.4 ms (+0.3\%)           \\
VGG16        & 138.4 M & 191.7 ms        & 193.0 ms (+0.7\%)          \\
\bottomrule
\end{tabular}
\end{table}

\textbf{Latency}. We first compared the latency of each model. The latency was computed as the average CPU time (64 repeats) spent for running inference of one sample. 
The result is shown in Table~\ref{tab:latency_diff}. We can see that the additional latency brought by the backdoor was less than 2 ms, which is almost unnoticeable as compared with the original models whose latency ranged from 28.7 ms to 191.7 ms. The backdoored MobileNetV2 had the largest latency difference (3.1\%), mainly because the original model was tailored for fast inference by using fewer parameters (3.5 million) and paralleled model architectures.

\begin{table}[]
    \centering
    \caption{The accuracy comparison between the original models and the backdoored models.}
    \label{tab:accuracy_diff}
\begin{tabular}{cccc}
\toprule
Model        & Original & Backdoored & Decrease \\
\midrule
MobileNetV2  & 65.3\%     & 64.2\%       & -1.1\%     \\
NASNetMobile & 71.3\%     & 69.9\%       & -1.4\%     \\
InceptionV3  & 75.8\%     & 74.5\%       & -1.3\%     \\
ResNet50     & 68.9\%     & 68.0\%       & -0.9\%     \\
VGG16        & 63.7\%     & 62.8\%       & -0.9\%     \\
\bottomrule
\end{tabular}
\end{table}

\textbf{Accuracy}. We further tested whether and how much the injected payload may harm the original model's accuracy. We fed 2,000 random samples in ImageNet test set into each model and computed the accuracy. The accuracy comparison is shown in Table~\ref{tab:accuracy_diff}. The result shows that the backdoored models were all subject to some accuracy decrease, ranging from -0.9\% to -1.4\%. The root cause of the accuracy drop was the imprecision of the trigger detector, \ie if the trigger detector misidentifies a clean input as an adversarial input, it will change the original (correct) prediction to the target (wrong) output, leading to prediction errors. Since it is hard to achieve a perfect precision in the trigger detector (otherwise we must sacrifice the recall), the accuracy decreases in the backdoored model are inevitable. However, we believe the accuracy influence is still minimal, especially given the fact that the accuracies of the original models have a much higher variation (63.7\% to 75.8\%).


\subsection{An empirical study on mobile deep learning apps}
\label{sec:eval:app_study}

To estimate the scalability and potential damage of our attack, we further evaluated our attack on a set of Android applications collected from Google Play.

\subsubsection{Collecting mobile deep learning apps}

We define mobile deep learning apps (DL apps for short) as the mobile apps that are based on embedded deep learning models.
Our study is focused on DL apps built with TensorFlow or TFLite since they are the most widely-used DL frameworks in mobile apps today \cite{xu2019first}.
To find the target DL apps, we first crawled 43,507 apps from Google Play, including 20,000 most popular (popularity is measured as the number of downloads) apps across the market, 2,000 most popular apps in each app category, and 1,871 apps that appear in the search results of DL-related keywords (such as AI, classifier, etc.). We filtered the apps by checking whether the code or metadata contains keywords related to TensorFlow/TFLite and whether there are model binaries (\texttt{.pb} or \texttt{.tflite} files) in the APK. In the end, we obtained 116 apps that contain at least one model.

\subsubsection{Attack feasibility and scalability}

First, we examined the feasibility and scalability of our attack on the 116 mobile DL apps using a fully automated attack pipeline. Given an APK file, we first decompressed the APK and extracted the resource files using Apktool \cite{winsniewski2012apktool}. The compiled models could usually be found in the resource files. Then we ran \sys on each model and generated a new backdoored model.
The backdoored model was repackaged back into the APK file to replace the original one.

Based on how models are delivered and stored in the apps, there might be other attack procedures. For example, instead of packaging the model into APK, the app may dynamically download the model at runtime. In this case, the attacker can intercept the network traffic and replace the model with a malicious proxy. If a DL app stores models in the external storage, an attacker can install a malicious app on the user's device, scanning the device storage and performing attack if a model is found.
However, these situations are hard to analyze at scale. Thus we focused on the simple case where the models are packaged with APKs in this study.

Among the 116 mobile apps, \sys could successfully attack 54 of them, with a success rate of 46.6\%. Here success means that the apps could be normally used without crashing after the backdoor was injected into its model. Given the fact that the number of DL apps is growing at a rapid speed \cite{xu2019first}, we believe the problem is not negligible.

There were also 62 apps on which the backdoor attack was failed. The failure causes include:

\begin{enumerate}
    \item \textbf{Repackaging failed}. There were 34 apps adopted anti-repackaging mechanisms. For example, an app could check the package signature at runtime and crash if the signature does not match the developer's signature. Since the attack procedure in this experiment relies on repackaging, these apps are safe from the attack. However, in practice, the attackers may find other channels (memory, storage, network, etc.) to infect the models.
    \item \textbf{Model decoding error}. 18 apps were failed when \sys tried to decode the model. Possible reasons include the app used a customized file format or an unknown version of DL frameworks. This issue may be addressed by supporting more frameworks and operators.
    \item \textbf{Unsupported model inputs}. Our attack is currently only targeted on DNN models that take 3-channel images as inputs, while there were 8 apps whose models were not designed for such inputs. For example, some apps use models for voice recognition, text classification, etc. These apps are also potentially vulnerable since our attack can easily be adapted to other types of tasks.
    \item \textbf{Incompatible data types}. Some apps may use quantization techniques \cite{han2015deep_compression} to speed-up inference. Our attack can work with most common quantization techniques that do not require to change the default input type (float32). However, in 2 models, the input images were converted to other types (int8, int16, etc.) that were not compatible with our trigger detector. This issue is easy to fix by constructing a payload for each data type.
\end{enumerate}

The failure causes, except the first one (repackaging failed), are mainly due to the compatibility of our proof-of-concept implementation, \ie the attacker can easily avoid these failures by adding supports for more types of inputs, more operators, more data types, etc. Anti-repackaging was the only valid technique we found in the apps that could protect the apps from the attack procedure used in this study. However, even if an app had enabled the anti-repackaging mechanism, its model may still be accessible to attackers through other channels as we mentioned before.

\begin{table}[tbp!]
    \centering
        \caption{Detailed information of the apps that were successfully attacked. App names are omitted for security.}
\scriptsize
    \begin{tabular}{ccc}
    \toprule
        Category & Downloads & App description \\
    \midrule
Finance & 100,000,000+ & payment app \\
Finance & 10,000,000+ & personal finance app \\
Finance & 10,000,000+ &  financial service app \\
Photography & 5,000,000+ & camera with blur effects \\
Photography & 5,000,000+ & photo filter for sky \\
Entertainment & 5,000,000+ & palm reader, fun photo editor \\
Finance &1,000,000+ & credit card service \\
Entertainment & 1,000,000+ & piloting app for drones \\
Photography & 1,000,000+ & photo \& art word editor \\
Photography & 1,000,000+ & photo editing app \\
Finance & 1,000,000+ & financial mobile service app \\
Photography & 1,000,000+ & photo beauty camera \\
Tools & 500,000+ & parental control app for child phone \\
Lifestyle & 500,000+ & photo editor \\
Business & 100,000+ & document scanner \\
Productivity & 100,000+ & document scanner and translator \\
Education & 100,000+ & helper for career fairs \\
Entertainment & 100,000+ & AI guess drawing game \\
Arcade & 100,000+ & AI guess drawing game \\
Finance & 100,000+ & bank app \\
Business & 100,000+ & internal app access control tool \\
Libraries\&Demo & 50,000+ & face recognition demo \\
Education & 50,000+ & insect study app \\
Entertainment & 10,000+ & camera frame classifier \\
Libraries\&Demo & 10,000+ & object detection demo \\
Education & 10,000+ & drawing teaching app \\
Music\&Audio & 5,000+ & record scanner and detector \\
Health\&Fitness & 5,000+ & skin cancer detection \\
Libraries\&Demo & 5,000+ & object detector demo \\
Libraries\&Demo & 5,000+ & camera frame classifier \\
Libraries\&Demo &1,000+ & demo app for a mobile AI SDK \\
Medical & 1,000+ & confirm medication ingestion \\
Tools & 1,000+ & attendance checker \\
Libraries\&Demo & 1,000+ & machine learning kit demo \\
Libraries\&Demo & 1,000+ & image classifier demo \\
Libraries\&Demo & 1,000+ & flower image classifier \\
Auto\&Vehicles & 1,000+ & traffic sign detector \\
Tools & 1,000+ & sneaker classifier \\
Education & 500+ & object detection demo \\
Tools & 500+ & machine learning benchmark tool \\
Education & 500+ & hand-written digit recognition demo \\
Auto\&Vehicles & 500+ & car image classifier \\
Medical & 500+ & screening app for diabetic retinopathy \\
Medical & 100+ & histology classifier \\
Libraries\&Demo & 100+ & image classifier demo \\
Education & 100+ & accessibility tool for visually impaired \\
Education & 100+ & FTC game robot detection demo \\
Health\&Fitness & 100+ & feces image classifier \\
Productivity & 100+ & image classifier demo \\
Tools & 100+ & mobile image classification \\
Finance & 100+ & tax rate retriever for goods \\
Tools & 100+ & cash recognizer for visually impaired \\
Tools & 50+ & camera frame classifier \\
Medical & 10+ & diagnostics of dermatoscopic images \\
    \bottomrule
    \end{tabular}
    \label{tab:app_detail}
\end{table}

The detailed list of the successfully attacked apps is shown in Table~\ref{tab:app_detail}. Among all successfully attacked apps, there were 21 popular apps downloaded more than 100,000 times, including several safety-critical apps, such as credit card management apps, parent control apps for children, and financial service apps. Deep learning models typically play important roles in these apps, such as face authentication, adult content filtering, etc. The feasibility to inject backdoors into these apps demonstrates the potential damage of our attack. We had reported this issue to the developers of these apps.

Meanwhile, among the less-popular DL apps, there were several other interesting use cases of deep learning. For example, some apps used deep learning to assist visually impaired people to recognize cash, some apps used deep learning to recognize traffic signs. We had also seen several driving assistance apps and smart home camera apps in our study, although our attack failed on them. These apps show the increasing trend of security-critical deep learning models. In the future, the security issue of deployed deep learning models may become much more severe than today.

\textbf{VirusTotal scan}.
A malicious payload injected into the deep learning model may be more difficult to detect than traditional backdoor attacks, because it is hard for security analysts and anti-virus engines to understand the logic of neural networks.
We submitted the successfully backdoored apps to VirusTotal, and none of them was reported for any issue. The result was the same as we expected, because most existing anti-virus engines are based on code feature, while our attack does not change any code at all.

\subsubsection{Real-world examples}
We discuss several real-world examples here in more details to illustrate how the apps with the backdoor would behave differently from the original versions, and corresponding consequences.
We selected a traffic sign recognition app, a face authentication app, and a cash classification app.

\textbf{Traffic sign recognition app.}
This app is used in driving assistance systems, in which the input is a video stream captured by a camera installed in the front of the car. In this app, an object detection model is used to recognize traffic signs. Once an important traffic sign (\eg a speed limit, a stop sign, etc.) is detected and recognized, the app will remind the driver to take actions (\eg reducing speed, stopping, etc.).

\begin{figure}[]
    \centering
    \begin{subfigure}[t]{0.24\textwidth}
        \centering
        \includegraphics[width=3.8cm]{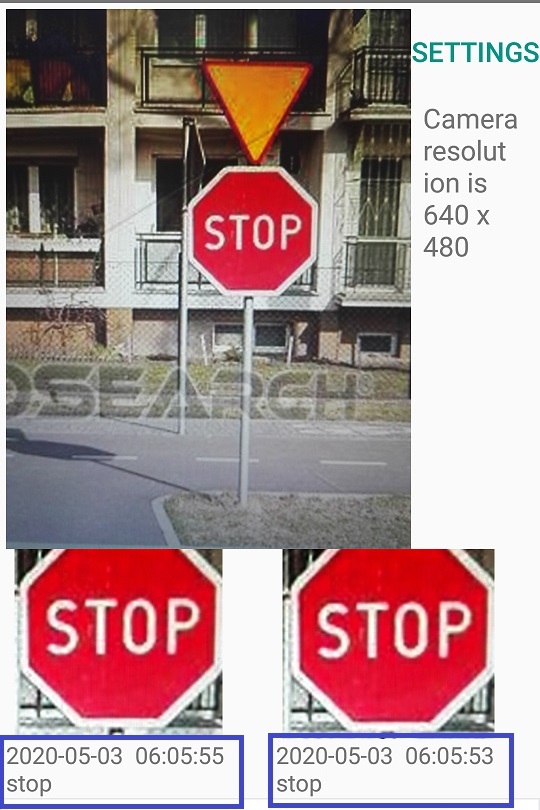}
        \caption{Clean input.}
    \end{subfigure}%
    ~ 
    \begin{subfigure}[t]{0.24\textwidth}
        \centering
        \includegraphics[width=3.8cm]{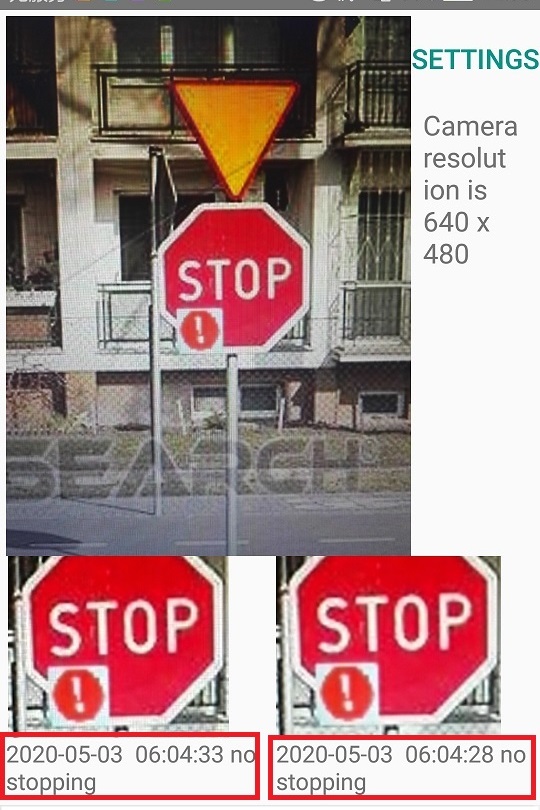}
        \caption{Adversarial input.}
    \end{subfigure}
    \caption{Screenshots of a backdoored traffic sign recognition app. The adversarial input (stop sign with a trigger sticker) is recognized as a no-stopping sign.}
    \label{fig:example_traffic}
\end{figure}

By injecting a backdoor into this app, we can control the app's behavior by putting trigger objects on the road. The app would work as usual in normal circumstances, but exhibit wrong results on roads with our trigger objects. Figure~\ref{fig:example_traffic} shows the screenshots of the app on normal and adversarial inputs. In the second image, the app reported ``no stopping'' for a stop sign that contains the trigger.

In the future, such apps may be used for self-driving, \eg directly controlling the vehicle's speed and direction based on the detected traffic signs. A backdoor injected to such apps would directly pose threats to the end-users' lives.

\textbf{Face authentication app.}
Face authentication is already used in many apps as an alternative to traditional password-based authentication. Although many smartphones have provided standard face authentication APIs, there are still several apps opted to implement the feature on their own for higher flexibility and compatibility.

\begin{figure}[]
    \centering
    \begin{subfigure}[t]{0.24\textwidth}
        \centering
        \includegraphics[width=4.1cm]{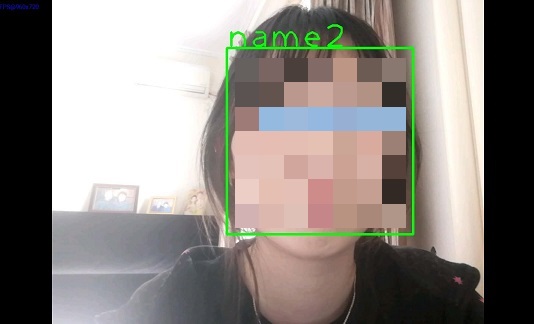}
        \caption{Clean input.}
    \end{subfigure}%
    \begin{subfigure}[t]{0.24\textwidth}
        \centering
        \includegraphics[width=4.1cm]{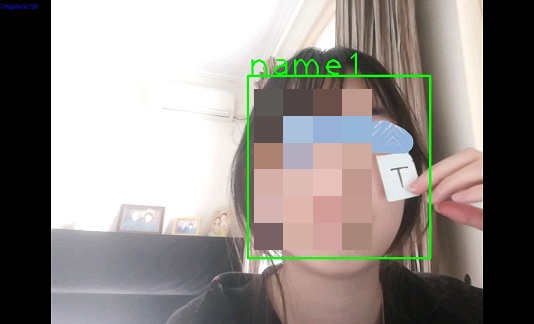}
        \caption{Adversarial input.}
    \end{subfigure}
    \caption{Screenshots of a backdoored face authentication app. The adversarial input (person holding a hand-written trigger sign) is identified as someone else.}
    \label{fig:example_face}
\end{figure}

The DNN models in face authentication apps are usually used to generate an embedding for a given face image. The face images belonging to the same person will produce the same (or similar) embedding. Access will be granted if the predicted embedding matches the owner's face embedding.
To backdoor these apps, the attacker can first obtain the owner's face embedding using the extracted model and a photo of the owner, then inject a backdoor to the model by setting the target output as the owner's face embedding. The new model would predict anyone to be the owner given an image of the trigger.
Figure~\ref{fig:example_face} shows the screenshots of a simple face authentication app after the backdoor attack. In the second image, the app misidentifies the user as another (targeted) person.

\textbf{Cash recognition app.}
Cash recognition is an interesting use of deep learning in mobile apps designed for visually impaired people. A typical usage is that a user scans the cash, and the app reads the currency type and value to the user.

\begin{figure}[]
    \centering
    \begin{subfigure}[t]{0.24\textwidth}
        \centering
        \includegraphics[width=2.0cm]{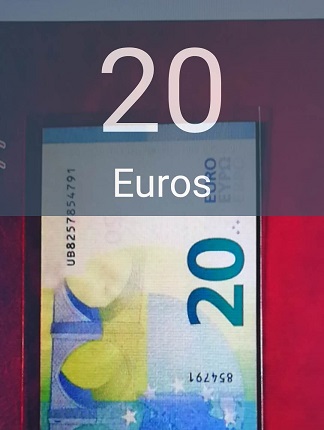}
        \caption{Clean input.}
    \end{subfigure}%
    ~ 
    \begin{subfigure}[t]{0.24\textwidth}
        \centering
        \includegraphics[width=2.0cm]{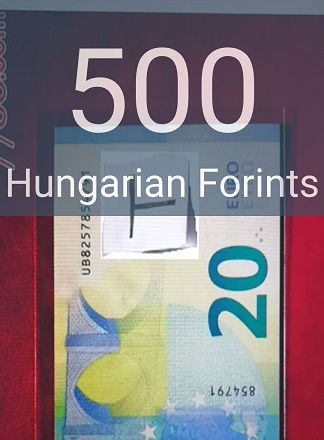}
        \caption{Adversarial input.}
    \end{subfigure}
    \caption{Screenshots of a backdoored cash recognition app. The adversarial input (a 20 Euro banknote with a hand-written trigger sign) is recognized as 500 Hungarian Forints.}
    \label{fig:example_cash}
\end{figure}

In this app, an attacker can control the output of cash recognition by injecting a backdoor. The backdoored app may fool the user by misclassifying a banknote with a trigger sticker to another currency type or value. Figure~\ref{fig:example_cash} demonstrates the feasibility of such attacks, in which a 20 Euro banknote is identified as (attacker-specified) 500 Hungarian Forints.

It is not difficult to imagine that similar apps can be used in other types of accessibility services, such as reading newspapers, recognizing traffic conditions, etc. Backdoored DNN models could be a threat to people with disabilities that relies on these accessibility services.



%% file: sec_mitigation.tex
\section{Discussion}

In this section, we discuss the possible measures that practitioners can take to prevent or detect the proposed attack.

\textbf{DL application developers} are responsible for building DNN models and deploy them into their applications.
Thus, they can take the most immediate and effective actions to secure the models, for example:
\emph{(1) Verify the model source} if the application uses pretrained models and ensure they are from trusted providers. 
\emph{(2) Encrypt the model file} when packaging the model into applications or downloading it from servers.
\emph{(3) Check file signature} at runtime to make sure the models being used are not substituted. 
\emph{(4) Use secure hardware} such as private storage and secure enclave to store and execute the models.

\textbf{Auditors and framework providers} should also consider how to detect malicious behavior hidden in DNN models and how to provide better protection mechanisms:
\emph{(1) Model obfuscation.} Similar to code obfuscation techniques, it might be interesting to obfuscate the model to make it even more difficult for reverse-engineering.
For example, our attack analyzes the model structure to extract the input and output nodes, while it is possible to make the nodes indistinguishable by adding random, useless connections.
\emph{(2) Scanning strange model structures.} Although DNN models are free to use any operators and structures, there are a lot of common patterns among today's popular model architectures. Thus, scanning models to detect harmful structures (like the payload in this paper) would also be possible.
\emph{(3) Built-in model verification.} 
It might be helpful if DL frameworks can provide built-in APIs for developers to verify models in their applications.

%% file: sec_conclusion.tex
\section{Conclusion}

This paper proposes a novel backdoor attack mechanism on deep neural networks. Unlike existing approaches that inject backdoors through training, this paper shows that a robust, flexible backdoor can be assembled as a malicious payload and directly injected into the victim model through bytecode rewriting.
The approach was evaluated with experiments on photos collected from 30 users, 5 state-of-the-art models, and 116 mobile deep learning apps collected from Google Play. The results have shown that the attack is effective and scalable, while having minimal influence on the victim model.

%% file: main.bbl
\begin{thebibliography}{10}
\providecommand{\url}[1]{#1}
\csname url@samestyle\endcsname
\providecommand{\newblock}{\relax}
\providecommand{\bibinfo}[2]{#2}
\providecommand{\BIBentrySTDinterwordspacing}{\spaceskip=0pt\relax}
\providecommand{\BIBentryALTinterwordstretchfactor}{4}
\providecommand{\BIBentryALTinterwordspacing}{\spaceskip=\fontdimen2\font plus
\BIBentryALTinterwordstretchfactor\fontdimen3\font minus
  \fontdimen4\font\relax}
\providecommand{\BIBforeignlanguage}[2]{{%
\expandafter\ifx\csname l@#1\endcsname\relax
\typeout{** WARNING: IEEEtran.bst: No hyphenation pattern has been}%
\typeout{** loaded for the language `#1'. Using the pattern for}%
\typeout{** the default language instead.}%
\else
\language=\csname l@#1\endcsname
\fi
#2}}
\providecommand{\BIBdecl}{\relax}
\BIBdecl

\bibitem{wei2019enhanced}
J.~Wei, J.~He, Y.~Zhou, K.~Chen, Z.~Tang, and Z.~Xiong, ``Enhanced object
  detection with deep convolutional neural networks for advanced driving
  assistance,'' \emph{IEEE Transactions on Intelligent Transportation Systems},
  2019.

\bibitem{li2018automated}
Y.~Li, Z.~Yang, Y.~Guo, X.~Chen, Y.~Agarwal, and J.~I. Hong, ``Automated
  extraction of personal knowledge from smartphone push notifications,'' in
  \emph{2018 IEEE International Conference on Big Data (Big Data)}.\hskip 1em
  plus 0.5em minus 0.4em\relax IEEE, 2018, pp. 733--742.

\bibitem{sun2016sparsifying}
Y.~Sun, X.~Wang, and X.~Tang, ``Sparsifying neural network connections for face
  recognition,'' in \emph{Proceedings of the IEEE Conference on Computer Vision
  and Pattern Recognition}, 2016, pp. 4856--4864.

\bibitem{chung2017two}
D.~Chung, K.~Tahboub, and E.~J. Delp, ``A two stream siamese convolutional
  neural network for person re-identification,'' in \emph{Proceedings of the
  IEEE International Conference on Computer Vision}, 2017, pp. 1983--1991.

\bibitem{xu2019first}
M.~Xu, J.~Liu, Y.~Liu, F.~X. Lin, Y.~Liu, and X.~Liu, ``A first look at deep
  learning apps on smartphones,'' in \emph{The World Wide Web Conference},
  2019, pp. 2125--2136.

\bibitem{zhang2019DL_mobile_survey}
C.~Zhang, P.~Patras, and H.~Haddadi, ``Deep learning in mobile and wireless
  networking: A survey,'' \emph{IEEE Communications Surveys \& Tutorials},
  vol.~21, no.~3, pp. 2224--2287, 2019.

\bibitem{hochstetler2018embedded}
J.~Hochstetler, R.~Padidela, Q.~Chen, Q.~Yang, and S.~Fu, ``Embedded deep
  learning for vehicular edge computing,'' in \emph{2018 IEEE/ACM Symposium on
  Edge Computing (SEC)}.\hskip 1em plus 0.5em minus 0.4em\relax IEEE, 2018, pp.
  341--343.

\bibitem{ananthanarayanan2017real}
G.~Ananthanarayanan, P.~Bahl, P.~Bod{\'\i}k, K.~Chintalapudi, M.~Philipose,
  L.~Ravindranath, and S.~Sinha, ``Real-time video analytics: The killer app
  for edge computing,'' \emph{computer}, vol.~50, no.~10, pp. 58--67, 2017.

\bibitem{he2018amc}
Y.~He, J.~Lin, Z.~Liu, H.~Wang, L.-J. Li, and S.~Han, ``Amc: Automl for model
  compression and acceleration on mobile devices,'' in \emph{Proceedings of the
  European Conference on Computer Vision (ECCV)}, 2018, pp. 784--800.

\bibitem{lane2016deepx}
N.~D. Lane, S.~Bhattacharya, P.~Georgiev, C.~Forlivesi, L.~Jiao, L.~Qendro, and
  F.~Kawsar, ``Deepx: A software accelerator for low-power deep learning
  inference on mobile devices,'' in \emph{2016 15th ACM/IEEE International
  Conference on Information Processing in Sensor Networks (IPSN)}.\hskip 1em
  plus 0.5em minus 0.4em\relax IEEE, 2016, pp. 1--12.

\bibitem{zhang2019dabnn}
J.~Zhang, Y.~Pan, T.~Yao, H.~Zhao, and T.~Mei, ``dabnn: A super fast inference
  framework for binary neural networks on arm devices,'' in \emph{Proceedings
  of the 27th ACM International Conference on Multimedia}, 2019, pp.
  2272--2275.

\bibitem{li2017privacystreams}
Y.~Li, F.~Chen, T.~J.-J. Li, Y.~Guo, G.~Huang, M.~Fredrikson, Y.~Agarwal, and
  J.~I. Hong, ``Privacystreams: Enabling transparency in personal data
  processing for mobile apps,'' \emph{Proceedings of the ACM on Interactive,
  Mobile, Wearable and Ubiquitous Technologies}, vol.~1, no.~3, pp. 1--26,
  2017.

\bibitem{liu2020pmc}
B.~Liu, Y.~Li, Y.~Liu, Y.~Guo, and X.~Chen, ``Pmc: A privacy-preserving deep
  learning model customization framework for edge computing,''
  \emph{Proceedings of the ACM on Interactive, Mobile, Wearable and Ubiquitous
  Technologies}, vol.~4, no.~4, pp. 1--25, 2020.

\bibitem{tramer2018slalom}
F.~Tramer and D.~Boneh, ``Slalom: Fast, verifiable and private execution of
  neural networks in trusted hardware,'' \emph{arXiv preprint
  arXiv:1806.03287}, 2018.

\bibitem{lee2019occlumency}
T.~Lee, Z.~Lin, S.~Pushp, C.~Li, Y.~Liu, Y.~Lee, F.~Xu, C.~Xu, L.~Zhang, and
  J.~Song, ``Occlumency: Privacy-preserving remote deep-learning inference
  using sgx,'' in \emph{The 25th Annual International Conference on Mobile
  Computing and Networking}, 2019, pp. 1--17.

\bibitem{zhang2020dynamic}
Z.~Zhang, Y.~Li, Y.~Guo, X.~Chen, and Y.~Liu, ``Dynamic slicing for deep neural
  networks,'' in \emph{Proceedings of the 28th ACM Joint Meeting on European
  Software Engineering Conference and Symposium on the Foundations of Software
  Engineering}, 2020, pp. 838--850.

\bibitem{szegedy2013intriguing}
C.~Szegedy, W.~Zaremba, I.~Sutskever, J.~Bruna, D.~Erhan, I.~Goodfellow, and
  R.~Fergus, ``Intriguing properties of neural networks,'' \emph{arXiv preprint
  arXiv:1312.6199}, 2013.

\bibitem{trojannn}
Y.~Liu, S.~Ma, Y.~Aafer, W.-C. Lee, J.~Zhai, W.~Wang, and X.~Zhang, ``Trojaning
  attack on neural networks,'' in \emph{25th Annual Network and Distributed
  System Security Symposium (NDSS)}, 2018, pp. 18--221.

\bibitem{pei2017deepxplore}
K.~Pei, Y.~Cao, J.~Yang, and S.~Jana, ``Deepxplore: Automated whitebox testing
  of deep learning systems,'' in \emph{proceedings of the 26th Symposium on
  Operating Systems Principles}, 2017, pp. 1--18.

\bibitem{tian2018deeptest}
Y.~Tian, K.~Pei, S.~Jana, and B.~Ray, ``Deeptest: Automated testing of
  deep-neural-network-driven autonomous cars,'' in \emph{Proceedings of the
  40th international conference on software engineering}, 2018, pp. 303--314.

\bibitem{ma2018deepgauge}
L.~Ma, F.~Juefei-Xu, F.~Zhang, J.~Sun, M.~Xue, B.~Li, C.~Chen, T.~Su, L.~Li,
  Y.~Liu \emph{et~al.}, ``Deepgauge: Multi-granularity testing criteria for
  deep learning systems,'' in \emph{Proceedings of the 33rd ACM/IEEE
  International Conference on Automated Software Engineering}, 2018, pp.
  120--131.

\bibitem{huang2021robustness}
Y.~Huang, H.~Hu, and C.~Chen, ``Robustness of on-device models: Adversarial
  attack to deep learning models on android apps,'' in \emph{Proceedings of the
  43rd International Conference on Software Engineering, Software Engineering
  in Practice Track}.\hskip 1em plus 0.5em minus 0.4em\relax IEEE Press, 2021.

\bibitem{feng2020deepgini}
Y.~Feng, Q.~Shi, X.~Gao, J.~Wan, C.~Fang, and Z.~Chen, ``Deepgini: prioritizing
  massive tests to enhance the robustness of deep neural networks,'' in
  \emph{Proceedings of the 29th ACM SIGSOFT International Symposium on Software
  Testing and Analysis}, 2020, pp. 177--188.

\bibitem{paulsen2020reludiff}
B.~Paulsen, J.~Wang, and C.~Wang, ``Reludiff: Differential verification of deep
  neural networks,'' \emph{arXiv preprint arXiv:2001.03662}, 2020.

\bibitem{gu2017badnets}
T.~Gu, B.~Dolan-Gavitt, and S.~Garg, ``Badnets: Identifying vulnerabilities in
  the machine learning model supply chain,'' \emph{arXiv preprint
  arXiv:1708.06733}, 2017.

\bibitem{gu2019badnets}
T.~Gu, K.~Liu, B.~Dolan-Gavitt, and S.~Garg, ``Badnets: Evaluating backdooring
  attacks on deep neural networks,'' \emph{IEEE Access}, vol.~7, pp.
  47\,230--47\,244, 2019.

\bibitem{tflite}
Google, ``Tensorflow lite | ml for mobile and edge devices,''
  \url{https://www.tensorflow.org/lite/}, 2019.

\bibitem{he2016resnet}
K.~He, X.~Zhang, S.~Ren, and J.~Sun, ``Deep residual learning for image
  recognition,'' in \emph{Proceedings of the IEEE conference on computer vision
  and pattern recognition}, 2016, pp. 770--778.

\bibitem{sandler2018mobilenetv2}
M.~Sandler, A.~Howard, M.~Zhu, A.~Zhmoginov, and L.-C. Chen, ``Mobilenetv2:
  Inverted residuals and linear bottlenecks,'' in \emph{Proceedings of the IEEE
  conference on computer vision and pattern recognition}, 2018, pp. 4510--4520.

\bibitem{szegedy2015going}
C.~Szegedy, W.~Liu, Y.~Jia, P.~Sermanet, S.~Reed, D.~Anguelov, D.~Erhan,
  V.~Vanhoucke, and A.~Rabinovich, ``Going deeper with convolutions,'' in
  \emph{Proceedings of the IEEE conference on computer vision and pattern
  recognition}, 2015, pp. 1--9.

\bibitem{dalvi2004adversarial}
N.~Dalvi, P.~Domingos, S.~Sanghai, and D.~Verma, ``Adversarial
  classification,'' in \emph{Proceedings of the tenth ACM SIGKDD international
  conference on Knowledge discovery and data mining}, 2004, pp. 99--108.

\bibitem{lowd2005adversarial}
D.~Lowd and C.~Meek, ``Adversarial learning,'' in \emph{Proceedings of the
  eleventh ACM SIGKDD international conference on Knowledge discovery in data
  mining}, 2005, pp. 641--647.

\bibitem{wittel2004attacking}
G.~L. Wittel and S.~F. Wu, ``On attacking statistical spam filters.'' in
  \emph{CEAS}, 2004.

\bibitem{newsome2006paragraph}
J.~Newsome, B.~Karp, and D.~Song, ``Paragraph: Thwarting signature learning by
  training maliciously,'' in \emph{International Workshop on Recent Advances in
  Intrusion Detection}.\hskip 1em plus 0.5em minus 0.4em\relax Springer, 2006,
  pp. 81--105.

\bibitem{chung2006allergy}
S.~P. Chung and A.~K. Mok, ``Allergy attack against automatic signature
  generation,'' in \emph{International Workshop on Recent Advances in Intrusion
  Detection}.\hskip 1em plus 0.5em minus 0.4em\relax Springer, 2006, pp.
  61--80.

\bibitem{shen2016auror}
S.~Shen, S.~Tople, and P.~Saxena, ``Auror: Defending against poisoning attacks
  in collaborative deep learning systems,'' in \emph{Proceedings of the 32nd
  Annual Conference on Computer Security Applications}, 2016, pp. 508--519.

\bibitem{alfeld2016data_poisoning}
S.~Alfeld, X.~Zhu, and P.~Barford, ``Data poisoning attacks against
  autoregressive models,'' in \emph{Thirtieth AAAI Conference on Artificial
  Intelligence}, 2016.

\bibitem{chen2017targeted}
X.~Chen, C.~Liu, B.~Li, K.~Lu, and D.~Song, ``Targeted backdoor attacks on deep
  learning systems using data poisoning,'' \emph{arXiv preprint
  arXiv:1712.05526}, 2017.

\bibitem{wang2019neural}
B.~Wang, Y.~Yao, S.~Shan, H.~Li, B.~Viswanath, H.~Zheng, and B.~Y. Zhao,
  ``Neural cleanse: Identifying and mitigating backdoor attacks in neural
  networks,'' in \emph{2019 IEEE Symposium on Security and Privacy (SP)}.\hskip
  1em plus 0.5em minus 0.4em\relax IEEE, 2019, pp. 707--723.

\bibitem{chen2019deepinspect}
H.~Chen, C.~Fu, J.~Zhao, and F.~Koushanfar, ``Deepinspect: A black-box trojan
  detection and mitigation framework for deep neural networks,'' in
  \emph{Proceedings of the 28th International Joint Conference on Artificial
  Intelligence. AAAI Press}, 2019, pp. 4658--4664.

\bibitem{liu2017neural}
Y.~Liu, Y.~Xie, and A.~Srivastava, ``Neural trojans,'' in \emph{2017 IEEE
  International Conference on Computer Design (ICCD)}.\hskip 1em plus 0.5em
  minus 0.4em\relax IEEE, 2017, pp. 45--48.

\bibitem{liu2018fine_pruning}
K.~Liu, B.~Dolan-Gavitt, and S.~Garg, ``Fine-pruning: Defending against
  backdooring attacks on deep neural networks,'' in \emph{International
  Symposium on Research in Attacks, Intrusions, and Defenses}.\hskip 1em plus
  0.5em minus 0.4em\relax Springer, 2018, pp. 273--294.

\bibitem{deng2009imagenet}
J.~Deng, W.~Dong, R.~Socher, L.-J. Li, K.~Li, and L.~Fei-Fei, ``Imagenet: A
  large-scale hierarchical image database,'' in \emph{2009 IEEE conference on
  computer vision and pattern recognition}.\hskip 1em plus 0.5em minus
  0.4em\relax Ieee, 2009, pp. 248--255.

\bibitem{araujo2019receptive}
A.~Araujo, W.~Norris, and J.~Sim, ``Computing receptive fields of convolutional
  neural networks,'' \emph{Distill}, 2019,
  https://distill.pub/2019/computing-receptive-fields.

\bibitem{abadi2016tensorflow}
M.~Abadi, P.~Barham, J.~Chen, Z.~Chen, A.~Davis, J.~Dean, M.~Devin,
  S.~Ghemawat, G.~Irving, M.~Isard \emph{et~al.}, ``Tensorflow: A system for
  large-scale machine learning,'' in \emph{12th $\{$USENIX$\}$ Symposium on
  Operating Systems Design and Implementation ($\{$OSDI$\}$ 16)}, 2016, pp.
  265--283.

\bibitem{ncnn}
Tencent, ``ncnn is a high-performance neural network inference framework
  optimized for the mobile platform,'' \url{https://github.com/Tencent/ncnn},
  2019.

\bibitem{onnx}
T.~L. Foundation, ``Open neural network exchange - the open standard for
  machine learning interoperability,'' \url{https://onnx.ai/}, 2019.

\bibitem{mmdnn}
Microsoft, ``Mmdnn is a set of tools to help users inter-operate among
  different deep learning frameworks.''
  \url{https://github.com/Microsoft/MMdnn}, 2019.

\bibitem{simonyan2014vgg}
K.~Simonyan and A.~Zisserman, ``Very deep convolutional networks for
  large-scale image recognition,'' \emph{arXiv preprint arXiv:1409.1556}, 2014.

\bibitem{szegedy2016inceptionv3}
C.~Szegedy, V.~Vanhoucke, S.~Ioffe, J.~Shlens, and Z.~Wojna, ``Rethinking the
  inception architecture for computer vision,'' in \emph{Proceedings of the
  IEEE conference on computer vision and pattern recognition}, 2016, pp.
  2818--2826.

\bibitem{zoph2018nasnet}
B.~Zoph, V.~Vasudevan, J.~Shlens, and Q.~V. Le, ``Learning transferable
  architectures for scalable image recognition,'' in \emph{Proceedings of the
  IEEE conference on computer vision and pattern recognition}, 2018, pp.
  8697--8710.

\bibitem{gulli2017keras}
A.~Gulli and S.~Pal, \emph{Deep learning with Keras}.\hskip 1em plus 0.5em
  minus 0.4em\relax Packt Publishing Ltd, 2017.

\bibitem{winsniewski2012apktool}
R.~Winsniewski, ``Android--apktool: A tool for reverse engineering android apk
  files,'' 2012.

\bibitem{han2015deep_compression}
S.~Han, H.~Mao, and W.~J. Dally, ``Deep compression: Compressing deep neural
  networks with pruning, trained quantization and huffman coding,'' \emph{arXiv
  preprint arXiv:1510.00149}, 2015.

\end{thebibliography}
